\documentclass[preprint,5p,times,twocolumn]{elsarticle}

\usepackage{amssymb}

\usepackage{lineno}

\usepackage{subcaption}
\usepackage{siunitx}
\DeclareSIUnit\GB{GB}
\DeclareSIUnit\arcmin{'}
\DeclareSIUnit\degsymbol{\degree}
\usepackage{xcolor}
\definecolor{darkorange}{RGB}{153, 76, 0}
\usepackage{arydshln}
\usepackage[hidelinks]{hyperref}
\def\code#1{\texttt{#1}}
\biboptions{numbers,sort&compress}

\journal{NIM-A}
\begin{document}
\begin{frontmatter}



\title{Signal-background separation and energy reconstruction of gamma rays using pattern spectra and convolutional neural networks for the Small-Sized Telescopes of the Cherenkov Telescope Array}


\author[inst1,inst2]{J. Aschersleben}
\author[inst1]{T. T. H. Arnesen}
\author[inst1]{R. F. Peletier}
\author[inst1]{M. Vecchi}
\author[inst1]{C. Vlasakidis}
\author[inst2]{M. H. F. Wilkinson}

\affiliation[inst1]{organization={Kapteyn Astronomical Institute, University of Groningen},
            addressline={PO Box 800}, 
            postcode={NL-9700 AV}, 
            city={Groningen},
            country={The Netherlands}}

\affiliation[inst2]{organization={Bernoulli Institute for Mathematics, Computer Science and Artificial Intelligence},
            addressline={PO Box 407}, 
            postcode={NL-9700 AK}, 
            city={Groningen},
            country={The Netherlands}}

\begin{abstract}
Imaging Atmospheric Cherenkov Telescopes (IACTs) detect very-high-energy gamma rays from ground level by capturing the Cherenkov light of the induced particle showers. Convolutional neural networks (CNNs) can be trained on IACT camera images of such events to differentiate the signal from the background and to reconstruct the energy of the initial gamma ray. Pattern spectra provide a 2-dimensional histogram of the sizes and shapes of features comprising an image and they can be used as an input for a CNN to significantly reduce the computational power required to train it. In this work, we generate pattern spectra from simulated gamma-ray and proton images to train a CNN for signal-background separation and energy reconstruction for the Small-Sized Telescopes (SSTs) of the Cherenkov Telescope Array (CTA). A comparison of our results with a CNN directly trained on CTA images shows that the pattern spectra-based analysis is about a factor of three less computationally expensive but not able to compete with the performance of an CTA image-based analysis. Thus, we conclude that the CTA images must be comprised of additional information not represented by the pattern spectra.

\end{abstract}


\end{frontmatter}

\section{Introduction} \noindent
When a gamma ray reaches the Earth's atmosphere, it induces a cascade of secondary particles which is known as an air shower. The secondary particles can reach velocities higher than the speed of light in air, inducing a flash of \textit{Cherenkov light}~\cite{Cherenkov1934}. The Cherenkov light can be captured by \textit{Imaging Air Cherenkov Telescopes} (IACTs) from the ground to reconstruct specific properties of the initial particle, such as its species, energy and direction (see~\cite{Sciascio_2019,de_Naurois_2015,De_Angelis_2018} for an overview of ground-based gamma-ray astronomy). 
The \textit{Cherenkov Telescope Array} (CTA)~\cite{CTA2019} is the next generation ground-based observatory for gamma-ray astronomy at very-high energies, offering 5-10 times better flux sensitivity than current generation gamma-ray telescopes~\cite{Gueta_2021}, such as H.E.S.S.~\cite{Benbow2006}, MAGIC~\cite{Bastieri2006} and VERITAS~\cite{Park2016}. It will cover a wide energy range between $\SI{20}{\GeV}$ to $\SI{300}{\TeV}$ benefiting from three different telescope types: \textit{Large-Sized Telescopes} (LSTs), \textit{Medium-Sized Telescopes} (MSTs) and \textit{Small-Sized Telescopes} (SSTs). The CTA Observatory will be distributed on two arrays in the northern hemisphere in La Palma (Spain) and the southern hemisphere near Paranal (Chile). CTA will outperform the energy and angular resolution of current instruments providing an energy resolution of $ \sim \SI{5}{\percent}$ around $\SI{1}{\TeV}$ and an angular resolution of $\SI{1}{\arcmin}$ at its upper energy range. With its short timescale capabilities and large field of view of $\SI{4.5}{\degsymbol}-\SI{8.5}{\degsymbol}$, it will enable the observation of a wide range of astronomical sources, including transient, high-variability or extended gamma-ray sources. \\
Several analysis methods for IACT data have been developed to classify the initial particle and reconstruct its energy and direction. \textit{Hillas parameters}~\cite{Hillas1985} are one of the first reconstruction techniques proposed by A. M. Hillas in 1985. They describe features of the Cherenkov emission within the camera images and are widely used as input to machine learning algorithms like \textit{Random Forest}~\cite{albert2008} or \textit{Boosted Decision Trees}~\cite{Ohm2009,Becherini2011,Krause2017} to perform full event reconstruction of gamma rays. Another approach is the \textit{ImPACT} algorithm~\cite{parsons2015hess}, which performs event reconstruction using expected image templates generated from Monte Carlo simulations. Other methods such as \textit{model analysis}~\cite{de_Naurois_2003} and \textit{3D model analysis}~\cite{LEMOINEGOUMARD2006195}, which are based on a semi-analytical shower model and a Gaussian photosphere shower model respectively, managed to be more sensitive to certain properties of the shower~\cite{Naurois2006}. \\
Recently, \textit{convolutional neural networks} (CNNs)~\cite{Gu2018,Wu2017,Li2021} have been proposed and applied to IACT data~\cite{feng2016analysis, Nieto2017, Mangano2018, Nieto2021a, Jacquemont2021, Aschersleben2021, Brill2019, Nieto2019, Miener2021b, Shilon2019, Miener2021a, Jacquemont2021b, spencer2021deep}. CNNs are machine learning algorithms that are specialised for image data and are currently one of the most successful tools for image classification and regression tasks~\cite{alzubaidi2021review}. They rely on \textit{convolutional layers} which consist of image filters that are able to extract relevant features within an image. Among many others, models such as \textit{AlexNet}~\cite{Krizhevsky2012}, \textit{GoogLeNet}~\cite{Szegedy2015} and \textit{ResNet}~\cite{He2016} established many new techniques, such as the \textit{Rectified Linear Unit} (ReLU)~\cite{Nair2010} activation function and deeper architectures, which set the milestones for many upcoming architectures. \textit{ResNet} won the \textit{ImageNet Large Scale Visual Recognition Challenge} (ILSVRC) in 2015 by introducing \textit{shortcut connections} into the architecture and achieving a \textit{top-5 classification error} of only $\SI{3.6}{\percent}$~\cite{ILSVRC15}. CNNs that contain these shortcut connections often achieve higher performances and are referred to as \textit{residual neural networks} (ResNets). The first event classifications with a CNN trained on IACT images have been presented in~\cite{feng2016analysis} and~\cite{Nieto2017}, which have demonstrated the signal-background separation capabilities of CNNs.
Later work has shown the energy and direction reconstruction capabilities of gamma rays with CNNs~\cite{Mangano2018, Nieto2021a, Jacquemont2021, Aschersleben2021}, their ability to run in stereo telescope mode~\cite{Brill2019, Nieto2019} and to be applied to real data~\cite{Shilon2019, Miener2021a, Jacquemont2021b}. 
In particular, the ResNet architecture has been shown to perform well for full event reconstruction for CTA data~\cite{Miener2021b}.
However, one of the main drawbacks of this method is that the training of CNNs is computationally very expensive~\cite{strubell2019energy}. It typically requires access to computing clusters with powerful graphics processing units (GPUs) and large amounts of random-access memory (RAM). The larger the dimension of the input image, the larger the computational power and time needed for the CNN training. A significant reduction of the dimension of the input image without any performance losses would therefore result in substantial savings in hardware and human resources, increase the efficiency of related scientific works and lower the environmental impact of CNNs~\cite{Lacoste2019}.
\\
An approach to this problem are \textit{pattern spectra}~\cite{Maragos1989}, which are commonly used tools for image classification~\cite{Urbach2007, batman1997size, chen1994gray} and can significantly reduce the computational power needed to train CNNs. They provide a 2-dimensional distribution of sizes and shapes of features within an image and can be constructed using a technique known as granulometries~\cite{Breen1996}. The features within the image are extracted with connected operators~\cite{salmbier2009}, which merge regions within an image with the same grey scale value. Compared to other feature extraction techniques, this approach has the advantage of not introducing any distortions into the image. In this work, we generate pattern spectra from simulated CTA images to apply them on a ResNet for signal-background separation and energy reconstruction of gamma rays. The application of a ResNet on pattern spectra takes advantage of their 2D nature by selecting relevant combinations of features within the CTA images. Our pattern spectra algorithm is based on the work presented in~\cite{Urbach2007}, which provides two main advantages compared to other existing pattern spectra algorithms: (i) the computing time for creating the pattern spectra is independent of its dimensions and (ii) it is significantly less sensitive to noise. These properties merit the investigation of pattern spectra-based analysis for IACTs. Direction reconstruction of gamma rays is not considered here since pattern spectra are rotation invariant, meaning that the same CTA image rotated by an arbitrary angle would result in the same pattern spectrum.
By generating pattern spectra from simulated CTA images, we aim to obtain a competitive algorithm that is significantly faster and less computationally intensive while keeping comparable performance to a CNN trained on CTA images in terms of signal-background separation and energy reconstruction of gamma rays. \\
The structure of this article is as follows: In Section~\ref{sec:dataset}, the CTA dataset used in this analysis is described. Section~\ref{sec:analysis} is devoted to our analysis methods including the pattern spectra algorithm, the ResNet architecture and the performance evaluation methods for our algorithms. The results are shown in Section~\ref{sec:results} and discussed in detail in Section~\ref{sec:discussion}. Finally, we state our conclusions in Section~\ref{sec:conclusion}. The source code of this project is publicly available at~\cite{jann_aschersleben_2023_8070256}.

\begin{figure}[ht]
    \centering
    \includegraphics[trim={0 2cm 0 2cm},clip]{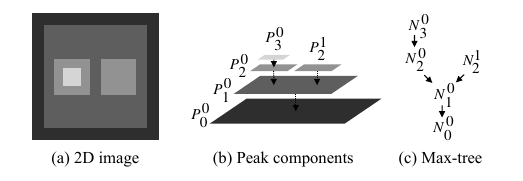}
    \caption{Visual representation of the pattern spectra algorithm (adapted from~\cite{Urbach2007,Teeninga2016})}
    \label{fig:max-tree}
\end{figure}
\begin{figure*}
    \centering
    \includegraphics[trim={0 0.7cm 0 0},clip,width=\textwidth]{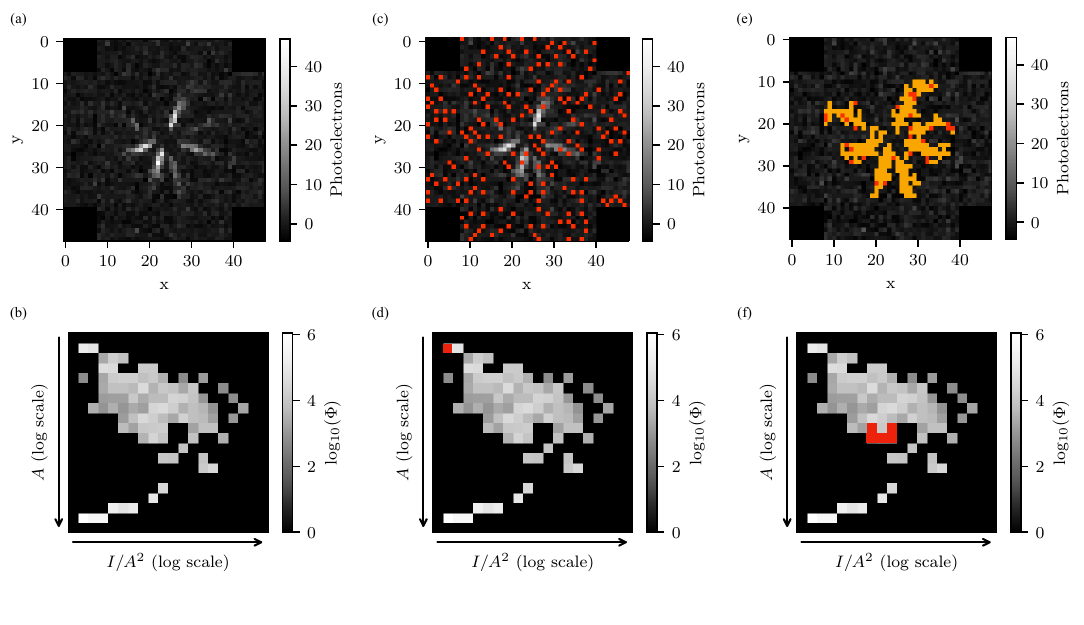}
    \caption{(a) CTA image of a $\SI{1.9}{\TeV}$ gamma-ray event captured by eight SSTs. (b) Pattern spectrum extracted from the CTA image. (c) CTA image with set of detected features highlighted in red. (d) Pattern spectrum with pixel corresponding to the detected features (small $A$ and $I/A^2$) highlighted in red. (e) CTA images with different set of detected (sub-)features highlighted in (red) orange. (f) Pattern spectrum with pixels corresponding to the detected features (intermediate $A$ and $I/A^2$) highlighted in red.}
    \label{fig:ps}
\end{figure*}

\begin{figure*}
    \centering
    \includegraphics[trim={0 0.7cm 0 0},clip,width=\textwidth]{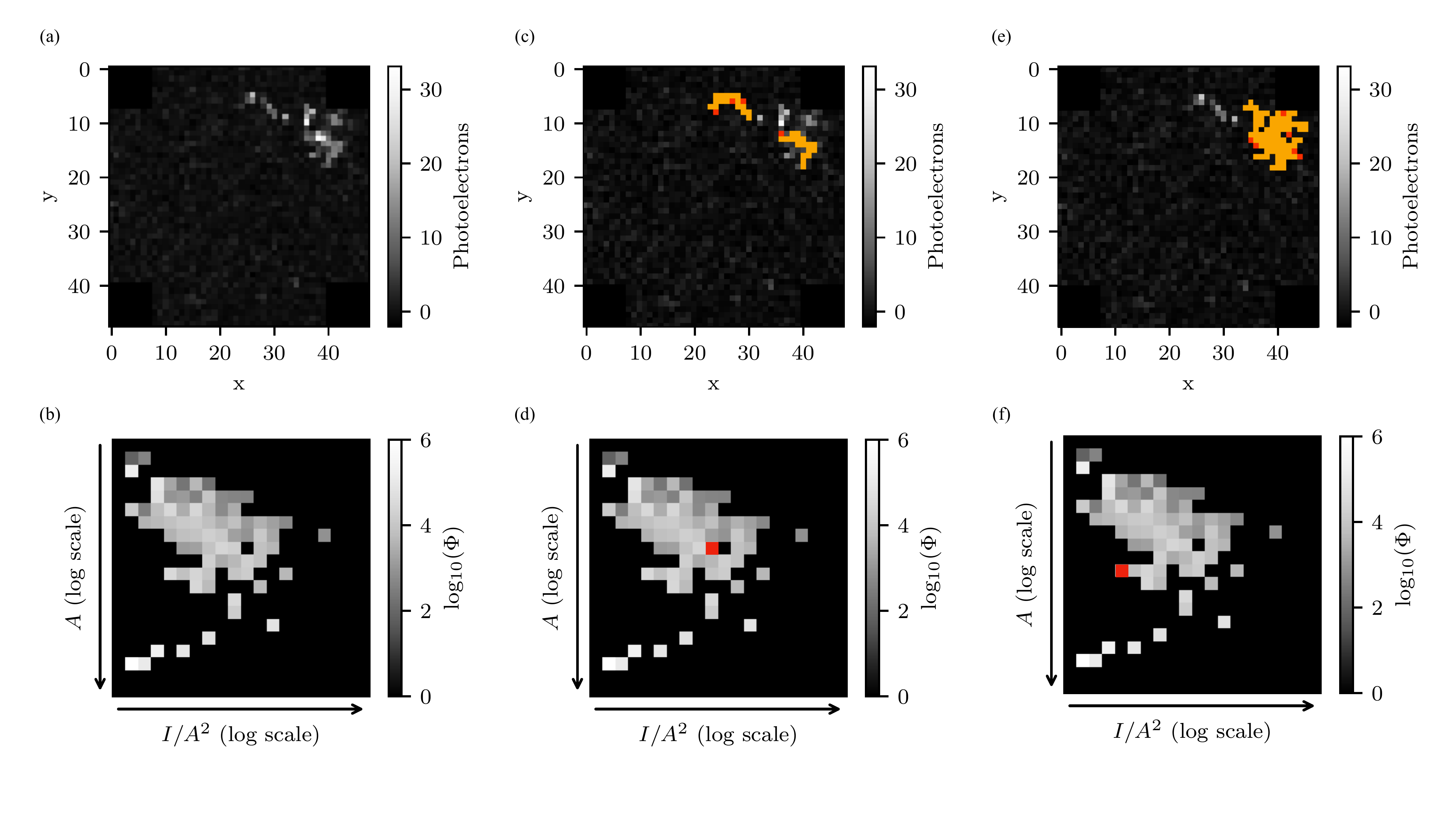}
    \caption{(a) CTA image of a $\SI{1.9}{\TeV}$ proton event. (b) Pattern spectrum extracted from the CTA image. (c) and (e): CTA image with a different set of detected features highlighted in orange and red. (d) and (f): Pattern spectrum with pixel corresponding to the detected features. Features with intermediate $A$ and intermediate $I/A^2$ are depicted in (c). Features with intermediate $A$ and small $I/A^2$ are depicted in (e).}
    \label{fig:ps_proton}
\end{figure*}

\begin{figure*}[ht]
    \centering
    \includegraphics{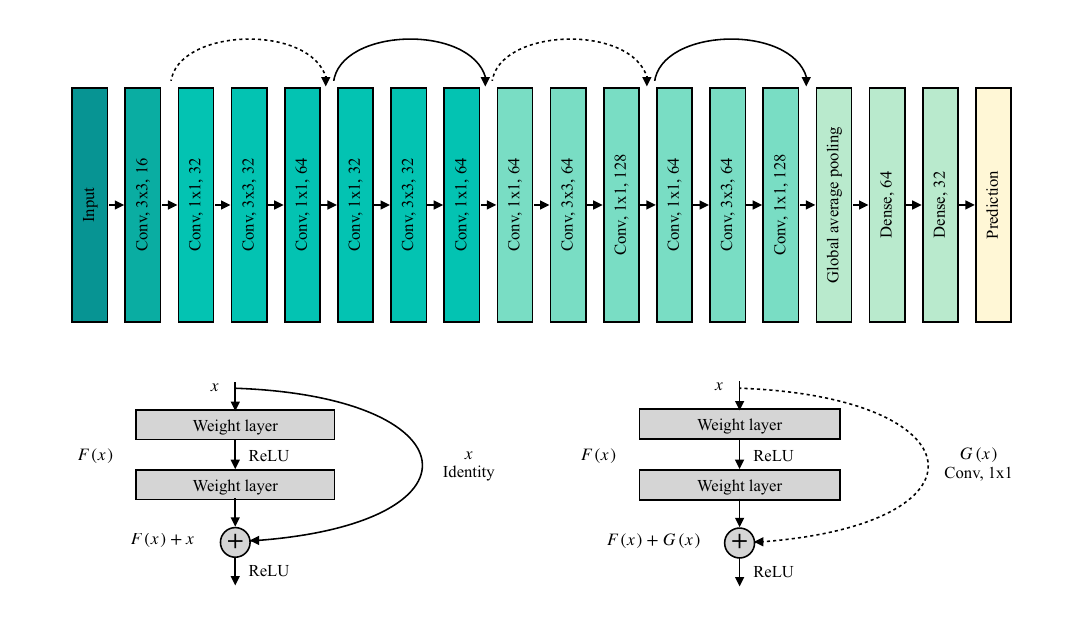}
    \caption{Top: Architecture of the thin residual neural network (TRN)~\cite{aschersleben2023event}. For each convolutional layer, the filter size and number of filters are specified. Bottom: Building block with a linear shortcut connection (left) and non-linear shortcut connection (right) (adapted from~\cite{He2016}).}
    \label{fig:trn}
\end{figure*}

\section{Dataset} 
\label{sec:dataset} \noindent
The dataset consists of simulated gamma-ray and proton events detected by the southern CTA array (\code{Prod5\_DL1 (ctapipe v0.10.5}~\cite{karl_kosack_2021_4581045}), zenith angle of $\SI{20}{\degsymbol}$, North pointing~\cite{bernlohr_konrad_2022_6218687, gueta2021cherenkov}). Due to the hexagonal pixels integrated in the LSTs and MSTs cameras, which cannot be processed by the current version of the pattern spectra algorithm, only the 37 SSTs with rectangular pixels are considered in this analysis. The SST images containing the charge information, i.e. the integrated photodetector pulse, will be referred to as \textit{CTA images} in the following. CTA images generated by gamma rays with an energy between $\SI{500}{\GeV}$ and $\SI{100}{\TeV}$ and protons with an energy between $\SI{1.5}{\TeV}$ and $\SI{100}{\TeV}$ have been considered for this study to match the operating energy range of the SSTs. \\
For the energy reconstruction $\sim 3 \cdot 10^6$ gamma-ray events generated with a $\SI{0.4}{\degsymbol}$ offset from the telescope pointing position, referred to as \textit{pointlike gamma rays} in the following, are used. For the signal-background separation $\sim 2 \cdot 10^6$ \textit{diffuse gamma rays} and $\sim 2 \cdot 10^6$ \textit{diffuse protons} are used, whereas the term \textit{diffuse} describes events generated in a view cone of $\SI{10}{\degsymbol}$. The pointlike and diffuse events are considered in the analysis to represent real observation conditions. When observing a source, background events reach the telescopes not only from the direction of the source but potentially from a much larger view cone. However, using pointlike gamma-rays and diffuse proton events for signal-background separation would introduce a bias for the learning process of the CNN. Therefore, we consider diffuse events for the signal-background separation and pointlike events for the energy reconstruction task. \\
In particular for high energies, the dataset often includes single events that were captured by multiple SSTs. This results in several CTA images for a single event. Since the construction and training of a CNN, that is able to handle a varying amount of input images, is very challenging, we constructed a single CTA image for each event as a first step towards the implementation of pattern spectra for the analysis of CTA images. In order to obtain a single CTA image per event, all CTA images of the same event are combined into a single image by adding up the individual pixel values of each image. We are aware that this is reducing the performance of the array, but we adopt this strategy to simplify our proof of concept work. Furthermore, we tested our analysis in mono-mode, i.e. using a single CTA image per event and found that the image stacking does not have any adverse effect on the performance of our pattern spectra analysis. However, we do not promote the idea of image stacking for CNN analyses with CTA data when trying to maximise the performance of the CNN. 

\begin{figure*}[htb]
    \centering
  
    \begin{subfigure}[b]{0.49\textwidth}
      \centering
      \includegraphics[width=\textwidth]{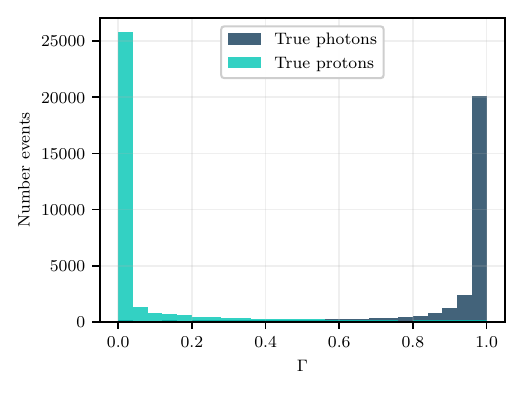}
    \end{subfigure}
    \hfill
    \begin{subfigure}[b]{0.49\textwidth}
      \centering
      \includegraphics[width=\textwidth]{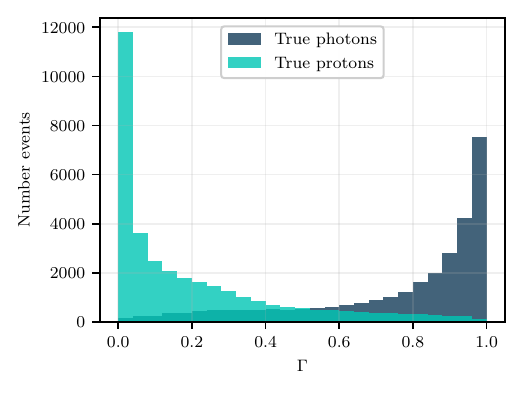}
    \end{subfigure}
  
    \caption{Example of the gammaness distributions obtained from a single TRN trained with CTA images (left) and pattern spectra (right).}
    \label{fig:gammaness}
\end{figure*}

\begin{figure*}[htb]
    \centering
  
    \begin{subfigure}[b]{0.49\textwidth}
      \centering
      \includegraphics[width=\textwidth]{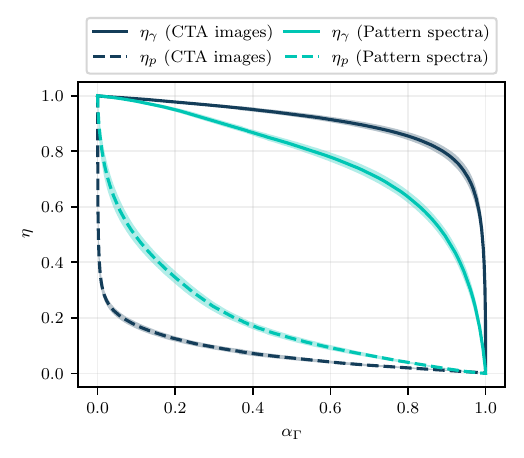}
    \end{subfigure}
    \hfill
    \begin{subfigure}[b]{0.49\textwidth}
      \centering
      \includegraphics[width=\textwidth]{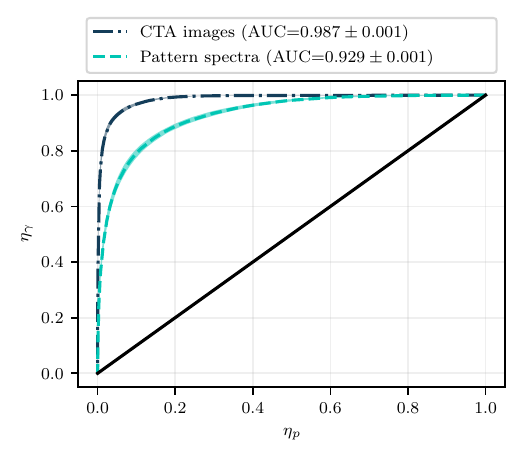}
    \end{subfigure}
  
    \caption{Left: Mean photon efficiency $\eta_{\gamma}$ and proton efficiency $\eta_{p}$ as a function of the $\Gamma$-threshold $\alpha_{\Gamma}$ obtained from 10 independent TRNs. Right: mean ROC curve and mean AUC-value obtained from 10 independent TRNs. The solid black line corresponds to a ROC curve expected from a random classifier. The performances stated here do not represent the expected performance by the CTA Observatory at the end of its construction phase.}
    \label{fig:efficiency_gammaness_roc}
\end{figure*}

\section{Analysis}
\label{sec:analysis}
\subsection{Pattern spectra} \noindent
The algorithm used to extract pattern spectra from the CTA images is based on the work presented in~\cite{Urbach2007} and will be briefly summarised in the following. \\
Let $f$ be a grey-scale image with grey levels $h$. 
In the case of CTA images, the grey levels $h$ correspond to the set of unique pixel values within the image. Consider an image domain $E \subseteq \mathbb{R}^2$ and let the set $X \subseteq E$ denote a binary image with domain $E$. The \textit{grain} of a binary image~$X$ is defined as a connected component $C$ of $X$. Therefore, grains are distinct regions that represent various structures and elements within the image~$X$. The \textit{peak components} $P^k_h(f)$ of an image~$f$ are defined as the $k$th grain of the threshold set~$T_h(f)$, which is defined as
\begin{equation}
    T_h(f) = \{x \in E | f(x) \geq h \}.
\end{equation}
Starting with the lowest grey level $h_0$ of the image $f$, the threshold set $T_{h_0}(f)$ always consists of a single peak component $P^{0}_{h_0}(f)$, independent of the image $f$. Increasing the grey level $h$ to the next larger value $h_1$, the threshold set $T_{h_1}(f)$ consists of $k$ peak components $P^{k}_{h_1}(f)$, which are the grains of the binary image $T_{h_1}(f)$. The grey levels are subsequently increased until the highest grey level $h_n$ is reached. Figure~\ref{fig:max-tree}~(a) shows an example of a 2D grey-scale image and (b) the corresponding peak components $P^k_h(f)$. In this particular example, the image consists of four grey levels $h = \{0,1,2,3\}$. For the grey levels  $h = \{0,1,3\}$, the threshold set $T_h(f)$ consists of a single peak component $P^{0}_{h_i}(f)$. For grey level $h=2$ two peak components, $P^{0}_{2}(f)$ and $P^{1}_{2}(f)$, are present. This is due to the fact that two distinct regions (grains) with grey level $h\geq2$ are present within the image. \\
Additionally to the threshold set $T_h(f)$, consider another set $Q_h(f)$ defined as
\begin{equation}
    Q_h(f) = \{x \in E | f(x) = h \}.
\end{equation}
The \textit{nodes} $N^k_h(f)$ of an image $f$ are defined as the connected components C of $T_h(f)$ such that $C \cap Q_h(f) \neq \emptyset$. A way to hierarchical represent the nodes $N^k_h(f)$ is the so called \textit{Max-tree}. The Max-tree of the previous example image is shown in Figure~\ref{fig:max-tree} (c). The root node of the Max-tree represents the set of pixels with the lowest grey level $h_0$, i.e. the set of pixels belonging to the background. The children of the root node represent the set of pixels with the next larger grey level $h_1$ and so on. The leaf nodes of the Max-tree represent the set of pixels with the highest grey level $h_n$, i.e. the set of pixels belonging to the foreground. For each image $f$, a Max-tree is computed according to the algorithm described in~\cite{Urbach2007}. \\
The pattern spectra are based on the size and shape attributes of the peak components $P^k_h(f)$. The size attribute corresponds to the area $A(P^k_h(f))$, which is computed by the sum of the pixels belonging to the detected feature. The shape attribute corresponds to $I/A^2$ with the moment of inertia $I$ describing the sum of squared differences to the centre of gravity of the feature. The size and shape attributes are binned into $N = 20$ size classes $s$ and shape classes $r$, which results in a good compromise between the performance of the pattern spectra and the computational power needed to train the ResNet. \\
The 2D pattern spectrum is computed from the Max-tree as follows~\cite{Urbach2007}:
\begin{enumerate}
    \item Construct a 2D array $\Phi[r,s]$ of size $N \times N = 20 \times 20$.
    \item Set all elements of $\Phi[r,s]$ to zero.
    \item For each node $N^k_h(f)$ of the Max-tree, compute the size class $r$ from the area $A(P^k_h(f))$, the shape class $s$ from $I(P^k_h(f))/A(P^k_h(f))^2$ and the grey-level difference $\delta_h$ between the current node and its parent.
    \item Add the product of $\delta_h$ and $A(P^k_h(f))$ to $\Phi[r,s]$.
\end{enumerate}
The current version of the algorithm is designed to work for images with square pixels but it could be adapted to work with hexagonal pixels in the future. \\
An example of a pattern spectrum extracted from a CTA image is shown in Figure~\ref{fig:ps}. 
The image in (a) shows a CTA image of a $\SI{1.9}{\TeV}$ gamma-ray event that was captured by eight SSTs. The bright features in the centre of the image correspond to the Cherenkov emission induced by the particle shower. Due to the different locations of the SSTs, the Cherenkov light is captured with different intensities and at different positions on the SST cameras. The pattern spectrum generated from the CTA image is shown in Figure~\ref{fig:ps} (b). Each pattern spectrum pixel represents a set of detected features. \\
An example of the detected features is shown in Figure~\ref{fig:ps} (c) \& (d). The image in (c) shows a set of detected features within the CTA image highlighted in red. The image in (d) shows the pattern spectrum with the red pixel representing these features. This specific example shows features with a small $A$ and small $I/A^2$ referring to features with a small size and a circular-like shape. They correspond to individual pixels in the CTA image and represent mostly noise. \\
Another example is shown in (e) and (f) of Figure~\ref{fig:ps}. Compared to the previous example, the red marked pattern spectrum pixels correspond to larger $A$ and $I/A^2$ values. Thus, the highlighted objects (red/orange) in the CTA image correspond to features with a larger size and more elliptical-like shape. The detected features in this example are of particular interest since they represent the Cherenkov photons induced by the particle shower, which contain information about the type and energy of the initial particle. \\
The pattern spectrum of a $\SI{1.9}{\TeV}$ proton event is shown in Figure~\ref{fig:ps_proton}. The features within the proton image differ significantly in comparison to the features present in the gamma-ray image shown in Figure~\ref{fig:ps}. Whereas the gamma-ray event results in mainly elliptical features, the features from the proton event vary notably more in shape and size. At a first glance, it is difficult to identify major differences between the pattern spectrum extracted from the gamma-ray event and the proton event. However, on closer inspection one can see that the gamma-ray pattern spectrum contains more features for larger $I/A^2$ and the proton pattern spectrum for smaller $I/A^2$ values. This meets our expectation because (i) gamma-ray events result in more elliptical-like features and (ii) the proton image contains more circular-like features. This does not mean, however, that proton events do not contain any elliptical-like features. Figure~\ref{fig:ps_proton} (c) and (e) show examples of the elliptical- and circular-like features for the proton event detected by the pattern spectrum algorithm. A classifier can therefore be trained on these differences to distinguish between gamma-ray and proton events.

\subsection{Residual neural network architecture} \noindent
For the signal-background separation and energy reconstruction of gamma-ray events, two individual but almost identical ResNet architectures are constructed and trained with either CTA images or pattern spectra. The architectures of our ResNets are identical to the ResNets presented in~\cite{aschersleben2023event} and are based on the work presented in~\cite{He2016,Xie2019,Miener2021b}. The ResNet is illustrated in Figure~\ref{fig:trn}. 
Due to the rather shallow architecture compared to the ResNet presented in~\cite{He2016}, we refer to our architectures as \textit{thin residual neural networks} (TRNs) in the following. They are constructed using \textit{Tensorflow 2.3.1}~\cite{Tensorflow_bib} and \textit{Keras 2.4.3}~\cite{Keras_bib} and consist of 13 convolutional layers with \textit{Rectified Linear Unit} (ReLU)~\cite{Nair2010} activation function, a \textit{global average pooling layer} and two \textit{fully connected (dense) layers} with 64 and 32 neurons respectively. The output layer consists of a single neuron for the energy reconstruction and two neurons with \textit{softmax}~\cite{Bridle1989} activation function for the signal-background separation. \textit{Shortcut connections}~\cite{He2016} at every third convolutional layer were implemented in order to improve the stability and performance of the algorithm. The solid arrows in Figure~\ref{fig:trn} represent linear shortcut connections, in which the input of a \textit{building block} $x$ is added to the output of the last layer of the building block $F(x)$. If the input and output of a building block have different dimensions, the input $x$ is put into another convolutional layer with the same number of filters as the last layer of the building block. The output of this residual operation $G(x)$ is added to the output of the last layer of the building block $F(x)$. A filter size of $1\times 1$ is used for all shortcut connections with a convolutional operation. In total, the two TRNs have about 150000 trainable parameters. 

\subsection{Network training and performance metrics} \noindent
The TRNs described in the previous section are trained and evaluated 10 times each on the datasets for both signal-background separation and energy reconstruction to perform a statistical analysis of the training process. Similar to the work presented in~\cite{Miener2021b}, a multiplicity cut of four or more triggered telescopes is applied for both the gamma-ray and proton events. The dataset is split into $\SI{90}{\percent}$ training data, from which $\SI{10}{\percent}$ is used as validation data, and $\SI{10}{\percent}$ test data. The weights of the TRN are initialized using the Glorot Uniform Initializer~\cite{glorot2010understanding} and the training, validation and test data are randomized for each run. The \textit{adaptive moment} (ADAM) optimizer~\cite{Kingma2014} with a learning rate of 0.001, a batch size of 32 is used for the TRN training. The training is stopped if there is no improvement on the validation dataset for over 20 epochs, and the model with the lowest validation loss is saved. The \textit{categorical cross entropy} and \textit{mean squared error}~\cite{Janocha2017} are applied as loss functions for the signal-background separation and energy reconstruction, respectively. The results shown in Section \ref{sec:results} are obtained by evaluating the performance of each TRN on the test data.

\begin{figure*}[htb]
    \centering
  
    \begin{subfigure}[b]{0.49\textwidth}
      \centering
      \includegraphics[width=\textwidth]{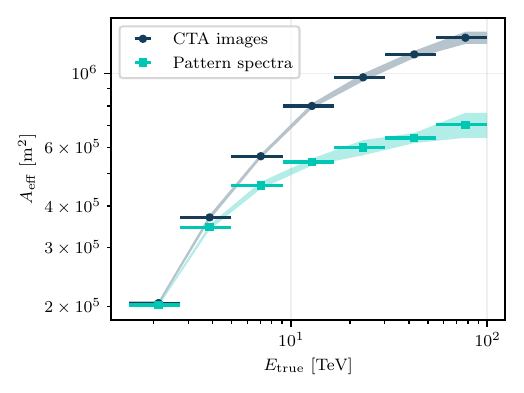}
    \end{subfigure}
    \hfill
    \begin{subfigure}[b]{0.49\textwidth}
      \centering
      \includegraphics[width=\textwidth]{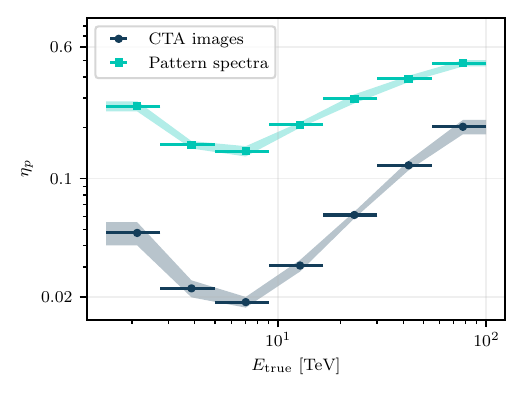}
    \end{subfigure}
  
    \caption{Left: mean effective area $A_{\text{eff}}$ as a function of the true energy $E_{\text{true}}$ obtained from 10 independent TRNs. Right: mean proton efficiency $\eta_p$ as a function of the true energy $E_{\text{true}}$ obtained from 10 independent TRNs. The proton efficiency was calculated by fixing the photon efficiency $\eta_\gamma$ to $\SI{90}{\percent}$ for each energy bin. Note that all simulated events are used to calculate the effective area $A_{\text{eff}}$ but only the events that pass the selection criteria are used to calculate the proton efficiency $\eta_p$ (see text for more details).}
    \label{fig:area_eff_eta_p}
\end{figure*}

\subsubsection{Signal-background separation} \noindent
Each event is labelled by its \textit{gammaness} $\Gamma$, whereas $\Gamma = 1$ corresponds to a gamma-ray (photon) and $\Gamma = 0$ corresponds to a proton. The output of the TRN is a $\Gamma$-value between 0 and 1, which describes a pseudo-probability of the event being a photon according to the TRN. For a fixed $\Gamma$-threshold $\alpha_{\Gamma}$, the \textit{photon efficiency} $\eta_{\gamma}$ is defined as $\eta_{\gamma} = TP / P$, where $TP$ is the number of \textit{true positives}, i.e. photon events with $\Gamma \geq \alpha_{\Gamma}$ (correctly classified photons), and $P$ is the total number of positives (photons) that pass the selection criteria described in Section~\ref{sec:dataset}.
Similarly, the \textit{proton efficiency} $\eta_{p}$ is defined as $\eta_{p} = FP / N$, where $FP$ is the number of \textit{false positives}, i.e. proton events with $\Gamma < \alpha_{\Gamma}$ (misclassified protons), and $N$ is the total number of negatives (protons) that pass the selection criteria. A good classifier results in a high photon efficiency $\eta_{\gamma}$ and a low proton efficiency $\eta_{p}$ for a given $\Gamma$-threshold. \\
In order to evaluate the performance of our TRNs, the efficiencies as a function of the $\Gamma$-threshold and the \textit{effective area} $A_{\text{eff}}$ as a function of the \textit{true energy} $E_{\text{true}}$ are calculated. The effectivate area is determined by $A_{\text{eff}} = \tilde{\eta}_{\gamma} \cdot A_{\text{geom}}$, where $A_{\text{geom}}$ is the geometrical area of the instrument, i.e. $A_{\text{geom}} = \pi r_{\text{max}}^2$ with $r_{\text{max}}$ being the maximum simulated impact radius, and $\tilde{\eta}_{\gamma}= TP / \tilde{P}$ with $\tilde{P}$ being the total number of simulated photons, including the events that did not pass the selection criteria in Section~\ref{sec:dataset}. Similarly, we define $\tilde{\eta}_{p} = FP / \tilde{N}$ with $\tilde{N}$ being the total number of simulated protons. 
The energy range is split into seven logarithmic bins, whereas each event is assigned to an energy bin based on its true energy $E_{\text{true}}$. The effective area is then calculated for each energy bin by increasing the $\Gamma$-threshold until $\tilde{\eta}_{p} = 10^{-3}$ is reached and extracting the corresponding $\tilde{\eta}_{\gamma}$. 
The value $\tilde{\eta}_{p} = 10^{-3}$ is motivated by the photon flux of the Crab Nebula being about three orders of magnitude lower than the isotropic flux of cosmic rays (CRs) within an angle of \SI{1}{\deg} around the direction of the source: $\Phi_{\gamma}^{\text{Crab}} \approx 10^{-3} \cdot \Phi_{\text{CR}}$ ~\cite{Sciascio_2019}.
Similarly, we determine the proton efficiency $\eta_p$ as a function of the true energy $E_{\text{true}}$ by fixing the photon efficiency $\eta_\gamma$ to $\SI{90}{\percent}$ for each energy bin.
Lastly, the \textit{receiver operating characteristic} (ROC) curve~\cite{BRADLEY1997} is determined. The ROC curve describes the photon efficiency $\eta_{\gamma}$ versus the proton efficiency $\eta_{p}$. The \textit{area under the ROC curve} (AUC) is calculated and used as a measure of the performance of each TRN. For part of our calculations we make use of \code{pyirf v0.7.0}~\cite{maximilian_linhoff_2023_7741289}, which is a \code{python} library for the generation of Instrument Response Functions (IRFs) and sensitivities for CTA. From the 10 TRNs, the mean efficiencies, effective area, ROC curve and the AUC-value are calculated for both the CTA images and pattern spectra-based analyses.

\subsection{Energy reconstruction} \noindent
The gamma-ray events are labelled by their true energy $E_{\text{true}}$, which the TRN learns to predict based on the training input. The performance of the TRN on the test data is evaluated by comparing the reconstructed energy $E_{\text{rec}}$ of the TRN with the true energy $E_{\text{true}}$ of the initial gamma ray. Therefore, the \textit{relative energy error} $\Delta E / E_{\text{true}} = (E_{\text{rec}} - E_{\text{true}}) / E_{\text{true}}$ is calculated for each event. The whole energy range between $\SI{500}{\GeV}$ and $\SI{100}{\TeV}$ is split into seven logarithmic bins and each event is assigned to an energy bin based on its true energy $E_{\text{true}}$. For each of these energy bins, the distribution of the relative energy error $\Delta E / E_{\text{true}}$ is determined and its median calculated. The median of $\Delta E / E_{\text{true}}$ is referred to as the \textit{energy bias} in the following. An energy bias close to zero indicates a good energy accuracy of the algorithm. The distributions of the relative energy error $\Delta E / E_{\text{true}}$ are then bias-corrected by subtracting the median, i.e. $(\Delta E / E_{\text{true}})_{\text{corr}} = \Delta E / E_{\text{true}} - \text{median}(\Delta E / E_{\text{true}})$. The \textit{energy resolution} is defined as the 68th percentile of the distribution $|(\Delta E / E_{\text{true}})_{\text{corr}}|$. From the 10 TRNs, the mean energy bias and energy resolution with their standard deviation are calculated for each energy bin for both the CTA images and pattern spectra-based analyses.

\begin{figure*}[htb]
    \centering
  
    \begin{subfigure}[b]{0.49\textwidth}
      \centering
      \includegraphics[width=\textwidth]{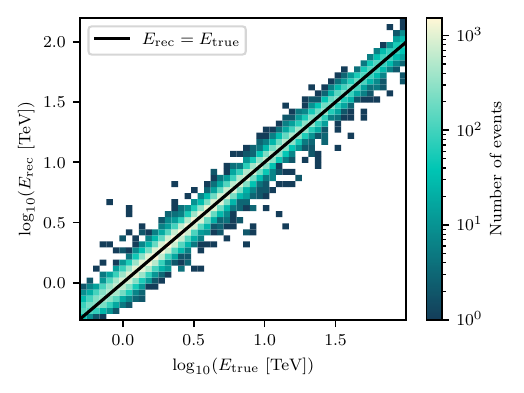}
    \end{subfigure}
    \hfill
    \begin{subfigure}[b]{0.49\textwidth}
      \centering
      \includegraphics[width=\textwidth]{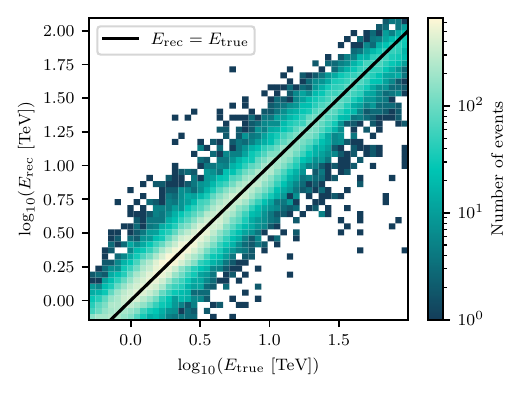}
    \end{subfigure}
  
    \caption{Example of the energy migration matrix obtained from a single TRN trained with CTA images (left) and pattern spectra (right).}
    \label{fig:energy_scattering_2D}
\end{figure*}

\section{Results}
\label{sec:results}

\subsection{Signal-background separation} \noindent
Two examples of the gammaness distributions obtained from a single TRN trained with the CTA images and pattern spectra are shown in Figure~\ref{fig:gammaness}. \\
Figure~\ref{fig:gammaness} (left) shows a distinct separation between photon and proton events for the TRN trained with CTA images. The majority of photon events are classified with $\Gamma = 1$ and the majority of proton events with $\Gamma = 0$. The number of proton (photon) events continuously decreases for larger (smaller) $\Gamma$-values, which indicates a good separation capability of the TRN. \\
Figure~\ref{fig:gammaness} (right) shows the performance of the TRN trained with the pattern spectra, which results in a lower signal-background separation capability compared to the TRN trained with CTA images. Once again, the majority of photon events are classified with $\Gamma = 1$ and the majority of proton events with $\Gamma = 0$. However, the distributions decrease less rapidly compared to the CTA images-based analysis. \\
The mean photon efficiency $\eta_{\gamma}$ and proton efficiency $\eta_{p}$ as a function of the $\Gamma$-threshold $\alpha_{\Gamma}$ are shown in Figure~\ref{fig:efficiency_gammaness_roc}. The shaded regions in this figure and the upcoming ones depict the standard deviation across the 10 TRNs. Both the photon efficiency and proton efficiency decrease steadily for an increasing $\alpha_{\Gamma}$-value. Up to $\Gamma \sim 0.1$ the pattern spectra-based analysis results in a very similar photon efficiency but in a much higher proton efficiency in comparison to the CTA images-based analysis. The proton efficiency of the pattern spectra approaches a similar value compared to the CTA images at $\Gamma \sim 0.9$ at which, however, the CTA images outperform the pattern spectra in the photon efficiency. Therefore, the CTA images results overall in a better photon and proton efficiencies independent of the $\Gamma$-threshold $\alpha_{\Gamma}$. \\
The mean ROC curve and corresponding AUC-value are shown in Figure~\ref{fig:efficiency_gammaness_roc} (right). As expected from the gammaness distributions discussed above, the ROC curve obtained from the CTA images is significantly steeper than the ROC curve obtained from the pattern spectra. The mean AUC-value of $0.987$ for the CTA images is therefore significantly larger than the value of $0.929$ obtained from the pattern spectra by a factor of $1.06$. \\
Figure~\ref{fig:area_eff_eta_p} (left) shows the mean effective area $A_{\text{eff}}$ as a function of the true energy $E_{\text{true}}$. The CTA images result in a higher effective area than the pattern spectra for all energies. The difference between the two analyses increases with increasing energy. The CTA images result in a maximum effective area of $\sim \SI{12.8e5}{\meter\squared}$ at $\sim \SI{80}{\TeV}$, whereas the pattern spectra result in a maximum effective area of $\sim \SI{7.0e5}{\meter\squared}$ at $\sim \SI{80}{\TeV}$, which corresponds to factor of 1.8 between the two analyses. \\
The mean proton efficiency $\eta_p$ as a function of the true energy $E_{\text{true}}$ is shown in Figure~\ref{fig:area_eff_eta_p} (right). The CTA images result in a lower proton efficiency, i.e. less misclassified protons, than the pattern spectra for all energies. For a fixed photon efficiency of $\SI{90}{\percent}$, both analyses achieve the lowest proton efficiency at $\sim \SI{7}{\TeV}$, whereas $\eta_p \approx \SI{2}{\percent}$ for the CTA images and $\eta_p \approx \SI{14}{\percent}$ for the pattern spectra. Percentage-wise, the difference is notably smaller for the higher energies. At the highest energy bin at $\sim \SI{80}{\TeV}$, the CTA images result in a proton efficiency of $\sim \SI{20}{\percent}$ and the pattern spectra in $\sim \SI{50}{\percent}$. \\
Overall, the TRN trained with CTA images shows a higher signal-background capability than the pattern spectra-based analysis. We discuss potential causes and implications of these results in Section~\ref{sec:discussion}.

\begin{figure*}[htb]
    \centering
  
    \begin{subfigure}[b]{0.49\textwidth}
      \centering
      \includegraphics[width=\textwidth]{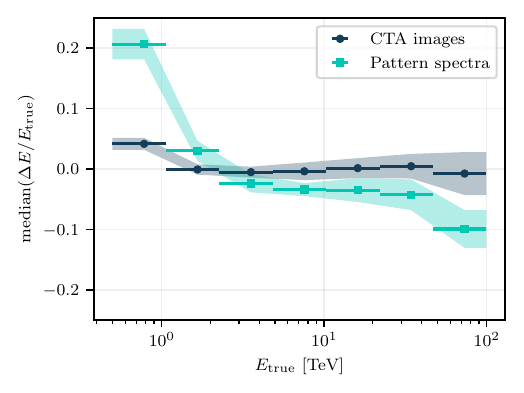}
    \end{subfigure}
    \hfill
    \begin{subfigure}[b]{0.49\textwidth}
      \centering
      \includegraphics[width=\textwidth]{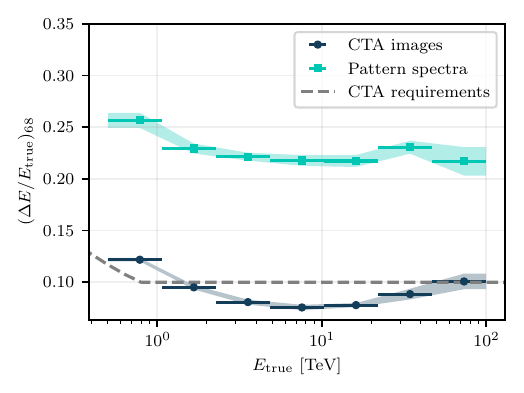}
    \end{subfigure}
  
    \caption{Mean energy accuracy (left) and resolution (right) obtained from 10 independent TRNs.}
    \label{fig:energy_res_acc}
\end{figure*}

\subsection{Energy reconstruction} \noindent
Figure~\ref{fig:energy_scattering_2D} shows two examples of the energy migration matrices, i.e. the 2D histogram of $E_{\text{rec}}$ against $E_{\text{true}}$, obtained from a single TRN trained with the CTA images and pattern spectra. \\
Most of the events are distributed around the $E_{\text{rec}} = E_{\text{true}}$ line for both the CTA images and pattern spectra-based analysis. However, the distribution obtained from the pattern spectra is more spread compared to the CTA images-based analysis. \\
The mean energy accuracy obtained from 10 independent TRNs is shown in Figure~\ref{fig:energy_res_acc} (left). The energy biases obtained from the CTA images-based analysis are closely distributed around $0$ with the largest energy bias of $\sim \SI{5}{\percent}$ at the lowest energy bin. The energy biases obtained from the pattern spectra-based analysis reaches up to $\sim \SI{20}{\percent}$ with the largest energy biases at the lowest and highest energy bin. The absolute value of the energy bias obtained from the pattern spectra-based analysis is larger than the values obtained from the CTA images for all energies. \\
The mean energy resolution obtained from 10 independent TRNs is shown in Figure~\ref{fig:energy_res_acc} (right). The CTA images-based analysis ranges from $0.08$ to $0.12$ with a minimum at $\sim \SI{7.5}{\TeV}$. While we simplified our analysis by stacking CTA images for each event, the energy resolution still meets the CTA requirements~\cite{cherenkov_telescope_array_observatory_2016_5163273} for all energy bins, except for the lowest energy bin. The pattern spectra result in an energy resolution between $0.22$ and $0.25$ with a minimum at the highest energy bin and does not meet the CTA requirements. Thus, the CTA images-based analysis outperforms the pattern spectra for all energies with a maximum factor of $2.9$ at $\sim \SI{7.5}{\TeV}$ between the two curves.

\begin{figure*}[htb]
    \centering
  
    \begin{subfigure}[b]{0.49\textwidth}
      \centering
      \includegraphics[width=\textwidth]{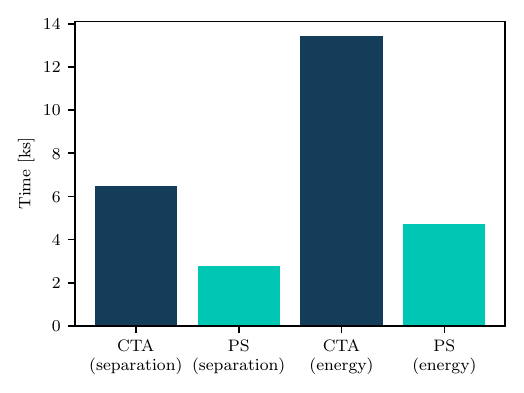}
    \end{subfigure}
    \hfill
    \begin{subfigure}[b]{0.49\textwidth}
      \centering
      \includegraphics[width=\textwidth]{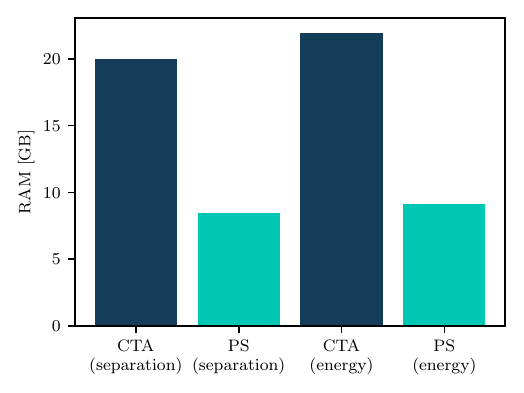}
    \end{subfigure}
  
    \caption{Mean time (left) and RAM (right) required to train the TRN for signal-background separation and energy reconstruction obtained from 10 independent TRNs for each analysis. The training was performed on a \textit{Nvidia A100 GPU}.}
    \label{fig:comp_power}
\end{figure*}

\section{Discussion} 
\label{sec:discussion} \noindent
A comparison of the computational performance of the analyses is shown in Figure~\ref{fig:comp_power}. The TRN training with pattern spectra is about a factor of 2.5 faster and requires a factor of 2.5 less RAM compared to the TRN training with CTA images. 
The pattern spectra are capable of detecting and classifying relevant features in the CTA images, which is illustrated by the gammaness distributions shown in Figure~\ref{fig:gammaness} (right) and the energy migration matrix shown in Figure~\ref{fig:energy_scattering_2D} (right). However, the pattern spectra-based analysis is outperformed by the CTA images with respect to their signal-background and energy reconstruction capabilities.
For a given $\Gamma$-threshold $\alpha_{\Gamma}$, the pattern spectra result in a poorer photon and proton efficiency compared to the CTA images (see Figure~\ref{fig:efficiency_gammaness_roc}), which is a main drawback of the analysis since both efficiencies are important quantities for the analysis of real gamma-ray data. Moreover, we infer from the effective area and proton efficiency versus energy plots shown in Figure~\ref{fig:area_eff_eta_p} that the signal-background capabilities of the pattern spectra-based analysis are below the capabilities of the CTA images-based analysis independent of the energy of the initial particle. Note that the different shape of the effective area and the proton efficiency curves are due to the fact that the effective area is determined using $\tilde{\eta}_\gamma$ and $\tilde{\eta}_p$, i.e. considering the total number of simulated photons and protons, whereas the proton efficiency $\eta_p$ in Figure~\ref{fig:area_eff_eta_p} (right) considers only those events that passed the selection criteria. 
The lowest proton efficiency $\eta_p$ is reached at an energy of $\sim \SI{5}{}-\SI{10}{\TeV}$, which corresponds to the energy range for which the SSTs are expected to have the highest flux sensitivity~\cite{cherenkov_telescope_array_observatory_2021_5499840}. For energies larger than $\sim \SI{10}{\TeV}$, we suppose that the increasing proton efficiency $\eta_p$ is caused by the increasing leakage, i.e. the fraction of the image intensity contained in the outermost pixels of the camera, in both the proton and gamma-ray images.
The AUC-value obtained from the CTA images is a factor 1.06 larger than the pattern spectra AUC-value and illustrates once again the overall lower signal-background capabilities of the pattern spectra-based analysis.
The CTA images result in a better energy resolution and a lower energy bias for all energies compared to the pattern spectra. Although our choice of attributes, i.e. size and shape attribute, is well-motivated, these two attributes do not seem to be sufficient to fully describe all relevant features within the CTA images. Potentially, the pattern spectra might not be able to detect, e.g., the electromagnetic substructure in proton showers. Other feature attributes, e.g. the \textit{perimeter}, \textit{sum of grey levels} and \textit{compactness} (\textit{perimeter} / $A^2$), were tested for both signal-background separation and energy reconstruction but did not result in a significantly better performance. 
Furthermore, we applied pattern spectra on other algorithms including classification and regression trees (CART)~\cite{Breiman1984}, Learning Vector Quantization (LVQ) and Generalized Matrix Learning Vector Quantization (GMLVQ)~\cite{Veen2021}. None of these algorithms achieved a better performance than the TRN. We therefore conclude that the TRN relies on features within the CTA images that are not detected by the pattern spectra algorithm. We suppose that the features within the CTA images are too complex to be sufficiently described by two attributes. For any given feature, one can always find a different feature with the same size and shape attribute values. This fact arguably makes it harder for any classifier to distinguish between gamma-ray and proton events or to reconstruct the energy of a gamma ray. Adding more than two attributes to the pattern spectrum, resulting in an n-dimensional pattern spectrum, might improve the performance of the algorithm. However, the computational power required to train a classifier with such an n-dimensional pattern spectra would significantly increase and would likely exceed the computational power required to train a classifier with CTA images. Given the results of the 2D pattern spectra presented in this work, we doubt that the n-dimensional pattern spectra would outperform the CTA images. Therefore, we decided to not pursue this idea further. \\
The performances stated in this work do not represent the expected performance by the CTA Observatory at the end of its construction phase.

\section{Conclusions} 
\label{sec:conclusion} \noindent
For the first time, signal-background separation and energy reconstruction of gamma rays was performed under the application of pattern spectra. We have shown that the pattern spectra algorithm has the capability to detect and classify relevant features in IACT images. The detected features are capable of differentiating between gamma-ray and proton events and to reconstruct the energy of gamma-ray events. The training of the TRN with pattern spectra requires 2.5 less RAM and is about a factor 2.5 faster than the TRN trained with CTA images, which agrees with our expectation due to the smaller size of the pattern spectra as compared to CTA images. The reduction in computational power was one of the main motivations to test the performance of pattern spectra on IACT data. However, the pattern spectra-based analysis is not competitive with the CTA images-based analysis in signal-background separation and energy reconstruction. The AUC-value, which is a measure of the signal-background separation capability of an algorithm, obtained from the CTA images is a factor 1.06 larger than the value obtained from the pattern spectra. The CTA images result in a better energy accuracy and energy resolution for all energies with a maximum factor of $2.9$ at $\sim \SI{7.5}{\TeV}$ in energy resolution compared to the pattern spectra. We, therefore, conclude that the relevant features within the CTA images are not sufficiently detected or described by our choice of size and shape attributes. Other sets of attributes were tested but resulted in no major improvements. Thus, the TRN trained on CTA images must rely on additional features not captured by the pattern spectra. In other applications, especially when the input images are larger, or vary in size, the results may be different.

\section*{Acknowledgements} \noindent
We extend our appreciation to Orel Gueta, Gernot Maier, Tjark Miener, Daniel Nieto-Castaño, Samuel Spencer and Thomas Vuillaume for their insightful discussions and valuable feedback on our work. Furthermore, we thank Deirdre Horan and Daniela Hadasch for managing the internal CTA reviewing process of this article. This work was conducted in the context of the CTA Consortium and CTA Observatory. We gratefully acknowledge financial support from the agencies and organizations listed at http://www.cta-observatory.org/consortium acknowledgements. We would like to thank the Center for Information Technology of the University of Groningen for their support and for providing access to the Peregrine high-performance computing cluster.

\section*{Declaration of generative AI and AI-assisted technologies in the writing process} \noindent
During the preparation of this work the authors used ChatGPT developed by OpenAI in order to improve language and readability of this manuscript. After using this tool/service, the authors reviewed and edited the content as needed and take full responsibility for the content of the publication.


 \bibliographystyle{elsarticle-num} 
 \bibliography{cas-refs}

\begin{thebibliography}{10}
\expandafter\ifx\csname url\endcsname\relax
  \def\url#1{\texttt{#1}}\fi
\expandafter\ifx\csname urlprefix\endcsname\relax\def\urlprefix{URL }\fi
\expandafter\ifx\csname href\endcsname\relax
  \def\href#1#2{#2} \def\path#1{#1}\fi

\bibitem{Cherenkov1934}
P.~A. Cherenkov, Visible emission of clean liquids by action of gamma
  radiation, Doklady Akad. Nauk SSSR 8~(451) (1934).

\bibitem{Sciascio_2019}
G.~D. Sciascio, Ground-based gamma-ray astronomy: an introduction, Journal of
  Physics: Conference Series 1263~(1) (2019) 012003.

\bibitem{de_Naurois_2015}
M.~de~Naurois, D.~Mazin, Ground-based detectors in very-high-energy gamma-ray
  astronomy, Comptes Rendus Physique 16~(6-7) (2015) 610--627.

\bibitem{De_Angelis_2018}
A.~D. Angelis, M.~Mallamaci, Gamma-ray astrophysics, The European Physical
  Journal Plus 133~(8) (2018).

\bibitem{CTA2019}
{CTA Consortium}, Science with the {Cherenkov Telescope Array}, WORLD
  SCIENTIFIC, 2019.

\bibitem{Gueta_2021}
O.~Gueta, {The {Cherenkov Telescope Array}: layout, design and performance},
  PoS ICRC2021 (2021) 885.

\bibitem{Benbow2006}
W.~Benbow, {HESS Collaboration}, The {HESS} experiment, in: Proceedings of AIP
  Conference, Vol. 842, American Institute of Physics, 2006, pp. 998--1000.

\bibitem{Bastieri2006}
D.~Bastieri, R.~Bavikadi, C.~Bigongiari, E.~Bisesi, P.~Boinee, A.~De~Angelis,
  B.~Lotto, A.~Forti, T.~Lenisa, F.~Longo, et~al., The {MAGIC} experiment and
  its first results, in: Frontiers of Fundamental Physics, Springer, 2006, pp.
  291--296.

\bibitem{Park2016}
N.~Park, Performance of the {VERITAS} experiment, in: Proceedings of the 34th
  International Cosmic Ray Conference, Vol. 236, SISSA Medialab, 2016, p. 771.

\bibitem{Hillas1985}
A.~M. Hillas, Cerenkov light images of {EAS} produced by primary gamma,
  International Cosmic Ray Conference 3 (1985).

\bibitem{albert2008}
J.~Albert, et~al., Implementation of the random forest method for the imaging
  atmospheric cherenkov telescope {MAGIC}, Nuclear Instruments and Methods in
  Physics Research Section A: Accelerators, Spectrometers, Detectors and
  Associated Equipment 588~(3) (2008) 424--432.

\bibitem{Ohm2009}
S.~Ohm, C.~{van Eldik}, K.~Egberts, gamma/hadron separation in very-high-energy
  gamma-ray astronomy using a multivariate analysis method, Astroparticle
  Physics 31~(5) (2009) 383--391.

\bibitem{Becherini2011}
Y.~Becherini, A.~Djannati-Ataï, V.~Marandon, M.~Punch, S.~Pita, A new analysis
  strategy for detection of faint gamma-ray sources with imaging atmospheric
  cherenkov telescopes, Astroparticle Physics 34~(12) (2011) 858--870.

\bibitem{Krause2017}
M.~Krause, E.~Pueschel, G.~Maier, Improved gamma/hadron separation for the
  detection of faint gamma-ray sources using boosted decision trees,
  Astroparticle Physics 89 (2017) 1--9.

\bibitem{parsons2015hess}
R.~Parsons, M.~Gajdus, T.~Murach, Hess ii data analysis with impact, arXiv
  preprint arXiv:1509.06322 (2015).

\bibitem{de_Naurois_2003}
M.~de~Naurois, J.~Guy, A.~Djannati-Atai, J.~P. Tavernet, Application of an
  analysis method based on a semi-analytical shower model to the first
  {H.E.S.S.} telescope, International Cosmic Ray Conference 28 ICRC (2003)
  2907--2910.

\bibitem{LEMOINEGOUMARD2006195}
M.~Lemoine-Goumard, B.~Degrange, M.~Tluczykont, Selection and
  {3D}-reconstruction of gamma-ray-induced air showers with a stereoscopic
  system of atmospheric cherenkov telescopes, Astroparticle Physics 25~(3)
  (2006) 195--211.

\bibitem{Naurois2006}
M.~de~Naurois, {Analysis methods for Atmospheric Cerenkov Telescopes}, in:
  Proceedings of 7th Workshop on Towards a Network of Atmospheric Cherenkov
  Detectors, 2005, pp. 149--162.

\bibitem{Gu2018}
J.~Gu, Z.~Wang, J.~Kuen, L.~Ma, A.~Shahroudy, B.~Shuai, T.~Liu, X.~Wang,
  G.~Wang, J.~Cai, T.~Chen, Recent advances in convolutional neural networks,
  Pattern Recognition 77 (2018) 354--377.

\bibitem{Wu2017}
J.~Wu, Introduction to convolutional neural networks, National Key Lab for
  Novel Software Technology. Nanjing University. China 5~(23) (2017) 495.

\bibitem{Li2021}
Z.~Li, F.~Liu, W.~Yang, S.~Peng, J.~Zhou, A survey of convolutional neural
  networks: Analysis, applications, and prospects, IEEE Transactions on Neural
  Networks and Learning Systems (2021) 1--21.

\bibitem{feng2016analysis}
Q.~Feng, T.~T. Lin, V.~Collaboration, et~al., The analysis of veritas muon
  images using convolutional neural networks, Proceedings of the International
  Astronomical Union 12~(S325) (2016) 173--179.

\bibitem{Nieto2017}
D.~Nieto~Castaño, A.~Brill, B.~Kim, T.~B. Humensky, {Exploring deep learning
  as an event classification method for the {Cherenkov Telescope Array}}, in:
  Proceedings of 35th International Cosmic Ray Conference {\textemdash}
  PoS(ICRC2017), Vol. 301, 2017, p. 809.

\bibitem{Mangano2018}
S.~Mangano, C.~Delgado, M.~I. Bernardos, M.~Lallena, J.~J.
  Rodr{\'\i}guez~V{\'a}zquez, {CTA Consortium}, et~al., Extracting gamma-ray
  information from images with convolutional neural network methods on
  simulated {Cherenkov Telescope Array} data, in: Proceedings of IAPR Workshop
  on Artificial Neural Networks in Pattern Recognition, Springer, 2018, pp.
  243--254.

\bibitem{Nieto2021a}
D.~Nieto, T.~Miener, A.~Brill, J.~Contreras, T.~Humensky, R.~Mukherjee,
  Reconstruction of {IACT} events using deep learning techniques with
  {CTLearn}, arXiv preprint, arXiv:2101.07626 (2021).

\bibitem{Jacquemont2021}
M.~Jacquemont, T.~Vuillaume, A.~Benoit, G.~Maurin, P.~Lambert, {Multi-Task
  Architecture with Attention for Imaging Atmospheric Cherenkov Telescope Data
  Analysis}, in: Proceedings of {VISAPP 2021}, 2021, p. none.

\bibitem{Aschersleben2021}
J.~Aschersleben, R.~F. Peletier, M.~Vecchi, M.~H.~F. Wilkinson, {Application of
  Pattern Spectra and Convolutional Neural Networks to the Analysis of
  Simulated {Cherenkov Telescope Array} Data}, in: Proceedings of 37th
  International Cosmic Ray Conference {\textemdash} PoS(ICRC2021), Vol. 395,
  2021, p. 697.

\bibitem{Brill2019}
A.~Brill, Q.~Feng, T.~B. Humensky, B.~Kim, D.~Nieto, T.~Miener, Investigating a
  deep learning method to analyze images from multiple gamma-ray telescopes,
  in: Proceedings of 2019 New York Scientific Data Summit (NYSDS), 2019, pp.
  1--4.

\bibitem{Nieto2019}
D.~Nieto, A.~Brill, Q.~Feng, T.~Humensky, B.~Kim, T.~Miener, R.~Mukherjee,
  J.~Sevilla, {CTLearn}: Deep learning for gamma-ray astronomy, arXiv preprint,
  arXiv:1912.09877 (2019).

\bibitem{Miener2021b}
T.~Miener, D.~Nieto, A.~Brill, S.~T. Spencer, J.~L. Contreras, {Reconstruction
  of stereoscopic {CTA} events using deep learning with {CTLearn}}, in:
  Proceedings of 37th International Cosmic Ray Conference {\textemdash}
  PoS(ICRC2021), Vol. 395, 2021, p. 730.

\bibitem{Shilon2019}
I.~Shilon, M.~Kraus, M.~Büchele, K.~Egberts, T.~Fischer, T.~Holch, T.~Lohse,
  U.~Schwanke, C.~Steppa, S.~Funk, Application of deep learning methods to
  analysis of imaging atmospheric cherenkov telescopes data, Astroparticle
  Physics 105 (2019) 44--53.

\bibitem{Miener2021a}
T.~Miener, R.~L{\'o}pez-Coto, J.~Contreras, J.~Green, D.~Green, E.~Mariotti,
  D.~Nieto, L.~Romanato, S.~Yadav, {IACT} event analysis with the {MAGIC}
  telescopes using deep convolutional neural networks with {CTLearn}, arXiv
  preprint, arXiv:2112.01828 (2021).

\bibitem{Jacquemont2021b}
M.~Jacquemont, T.~Vuillaume, A.~Benoit, G.~Maurin, P.~Lambert, G.~Lamanna,
  First full-event reconstruction from imaging atmospheric cherenkov telescope
  real data with deep learning, in: Proceedings of 2021 International
  Conference on Content-Based Multimedia Indexing (CBMI), 2021, pp. 1--6.

\bibitem{spencer2021deep}
S.~Spencer, T.~Armstrong, J.~Watson, S.~Mangano, Y.~Renier, G.~Cotter, Deep
  learning with photosensor timing information as a background rejection method
  for the {Cherenkov Telescope Array}, Astroparticle Physics 129 (2021) 102579.

\bibitem{alzubaidi2021review}
L.~Alzubaidi, J.~Zhang, A.~J. Humaidi, A.~Al-Dujaili, Y.~Duan, O.~Al-Shamma,
  J.~Santamar{\'\i}a, M.~A. Fadhel, M.~Al-Amidie, L.~Farhan, Review of deep
  learning: Concepts, cnn architectures, challenges, applications, future
  directions, Journal of big Data 8~(1) (2021) 1--74.

\bibitem{Krizhevsky2012}
A.~Krizhevsky, I.~Sutskever, G.~E. Hinton, Imagenet classification with deep
  convolutional neural networks, Advances in neural information processing
  systems 25 (2012).

\bibitem{Szegedy2015}
C.~Szegedy, W.~Liu, Y.~Jia, P.~Sermanet, S.~Reed, D.~Anguelov, D.~Erhan,
  V.~Vanhoucke, A.~Rabinovich, Going deeper with convolutions, in: Proceedings
  of the IEEE conference on computer vision and pattern recognition, 2015, pp.
  1--9.

\bibitem{He2016}
K.~He, X.~Zhang, S.~Ren, J.~Sun, Deep residual learning for image recognition,
  in: Proceedings of the IEEE conference on computer vision and pattern
  recognition, 2016, pp. 770--778.

\bibitem{Nair2010}
V.~Nair, G.~E. Hinton, Rectified linear units improve restricted boltzmann
  machines, in: Proceedings of the 27th International Conference on
  International Conference on Machine Learning, ICML'10, Omnipress, Madison,
  WI, USA, 2010, pp. 807--814.

\bibitem{ILSVRC15}
O.~Russakovsky, J.~Deng, H.~Su, J.~Krause, S.~Satheesh, S.~Ma, Z.~Huang,
  A.~Karpathy, A.~Khosla, M.~Bernstein, A.~C. Berg, L.~Fei-Fei, {ImageNet Large
  Scale Visual Recognition Challenge}, International Journal of Computer Vision
  (IJCV) 115~(3) (2015) 211--252.

\bibitem{strubell2019energy}
E.~Strubell, A.~Ganesh, A.~McCallum, Energy and policy considerations for deep
  learning in nlp, arXiv preprint arXiv:1906.02243 (2019).

\bibitem{Lacoste2019}
A.~Lacoste, A.~Luccioni, V.~Schmidt, T.~Dandres, Quantifying the carbon
  emissions of machine learning, arXiv preprint, arXiv:1910.09700 (2019).

\bibitem{Maragos1989}
P.~{Maragos}, Pattern spectrum and multiscale shape representation, IEEE
  Transactions on Pattern Analysis and Machine Intelligence 11~(7) (1989)
  701--716.

\bibitem{Urbach2007}
E.~R. Urbach, J.~B. T.~M. Roerdink, M.~H.~F. Wilkinson, Connected shape-size
  pattern spectra for rotation and scale-invariant classification of gray-scale
  images, IEEE transactions on pattern analysis and machine intelligence
  (2007).

\bibitem{batman1997size}
S.~Batman, E.~R. Dougherty, Size distributions for multivariate morphological
  granulometries: texture classification and statistical properties, Optical
  Engineering 36~(5) (1997) 1518--1529.

\bibitem{chen1994gray}
Y.~Chen, E.~R. Dougherty, Gray-scale morphological granulometric texture
  classification, Optical Engineering 33~(8) (1994) 2713--2722.

\bibitem{Breen1996}
E.~J. Breen, R.~Jones, Attribute openings, thinnings, and granulometries,
  Computer Vision and Image Understanding 64~(3) (1996) 377--389.

\bibitem{salmbier2009}
P.~Salembier, M.~H.~F. Wilkinson, Connected operators, Signal Processing
  Magazine, IEEE 26 (2009) 136 -- 157.

\bibitem{jann_aschersleben_2023_8070256}
J.~Aschersleben, T.~T.~H. Arnesen, R.~F. Peletier, M.~Vecchi, M.~H.~F.
  Wilkinson, \href{https://doi.org/10.5281/zenodo.8070256}{jaschers/psnet:
  v1.0} (Jun. 2023).
\newblock \href {https://doi.org/10.5281/zenodo.8070256}
  {\path{doi:10.5281/zenodo.8070256}}.
\newline\urlprefix\url{https://doi.org/10.5281/zenodo.8070256}

\bibitem{Teeninga2016}
P.~Teeninga, U.~Moschini, S.~C. Trager, M.~H.~F. Wilkinson, Statistical
  attribute filtering to detect faint extended astronomical sources,
  Mathematical Morphology - Theory and Applications 1~(1) (2016).

\bibitem{aschersleben2023event}
J.~Aschersleben, M.~Vecchi, M.~H.~F. Wilkinson, R.~F. Peletier, Event
  reconstruction using pattern spectra and convolutional neural networks for
  the {Cherenkov Telescope Array}, arXiv preprint arXiv:2302.11876 (2023).

\bibitem{karl_kosack_2021_4581045}
K.~Kosack, J.~Watson, M.~Nöthe, J.~Jacquemier, A.~Mitchell, D.~Neise, C.~Deil,
  S.~T. Spencer, R.~de~los Reyes, F.~Cassol, K.~Brügge, M.~Mastropietro,
  T.~Vuillaume, J.~Decock, moralejo, W.~Bhattacharyya, francesco visconti,
  L.~Nickel, J.~E. Ruiz, mgaug, N.~Biederbeck, M.~Peresano, R.~Lopez-Coto,
  orelgueta, C.~Alispach, J.~Lefaucheur, K.~Pfrang, M.~Hütten,
  thomasarmstrong, A.~Donini,
  \href{https://doi.org/10.5281/zenodo.4581045}{cta-observatory/ctapipe:
  v0.10.5} (Mar. 2021).
\newblock \href {https://doi.org/10.5281/zenodo.4581045}
  {\path{doi:10.5281/zenodo.4581045}}.
\newline\urlprefix\url{https://doi.org/10.5281/zenodo.4581045}

\bibitem{bernlohr_konrad_2022_6218687}
K.~Bernlöhr, O.~Gueta, G.~Maier, A.~Moralejo, Y.~Suda,
  \href{https://doi.org/10.5281/zenodo.6218687}{{CTAO Simulation Telescope
  Models for CORSIKA and sim\_telarray - prod5}} (Feb. 2022).
\newblock \href {https://doi.org/10.5281/zenodo.6218687}
  {\path{doi:10.5281/zenodo.6218687}}.
\newline\urlprefix\url{https://doi.org/10.5281/zenodo.6218687}

\bibitem{gueta2021cherenkov}
O.~Gueta, The cherenkov telescope array: layout, design and performance, arXiv
  preprint arXiv:2108.04512 (2021).

\bibitem{Xie2019}
W.~Xie, A.~Nagrani, J.~S. Chung, A.~Zisserman, Utterance-level aggregation for
  speaker recognition in the wild, in: Proceedings of ICASSP 2019-2019 IEEE
  International Conference on Acoustics, Speech and Signal Processing (ICASSP),
  IEEE, 2019, pp. 5791--5795.

\bibitem{Tensorflow_bib}
{M. Abadi {\em et al.}}, Tensorflow: Large-scale machine learning on
  heterogeneous systems, available at \url{tensorflow.org}, accessed on
  September 14th, 2022 (2015).

\bibitem{Keras_bib}
{F. Chollet {\em et al.}}, Keras, available at
  \url{https://github.com/fchollet/keras}, accessed on September 14th, 2022
  (2015).

\bibitem{Bridle1989}
J.~S. Bridle, Training stochastic model recognition algorithms as networks can
  lead to maximum mutual information estimation of parameters, in: Proceedings
  of the 2nd International Conference on Neural Information Processing Systems,
  NIPS'89, MIT Press, Cambridge, MA, USA, 1989, pp. 211--217.

\bibitem{glorot2010understanding}
X.~Glorot, Y.~Bengio, Understanding the difficulty of training deep feedforward
  neural networks, in: Proceedings of the thirteenth international conference
  on artificial intelligence and statistics, JMLR Workshop and Conference
  Proceedings, 2010, pp. 249--256.

\bibitem{Kingma2014}
D.~P. Kingma, J.~Ba, {Adam}: A method for stochastic optimization, arXiv
  preprint, arXiv:1412.6980 (2014).

\bibitem{Janocha2017}
K.~Janocha, W.~M. Czarnecki, On loss functions for deep neural networks in
  classification, arXiv preprint, arXiv:1702.05659 (2017).

\bibitem{BRADLEY1997}
A.~P. Bradley, The use of the area under the {ROC} curve in the evaluation of
  machine learning algorithms, Pattern Recognition 30~(7) (1997) 1145--1159.

\bibitem{maximilian_linhoff_2023_7741289}
M.~Linhoff, M.~Peresano, R.~M. Dominik, J.~Sitarek, T.~Vuillaume, M.~Punch,
  L.~Nickel, N.~Biederbeck, G.~Maier, A.~Moralejo, L.~Jouvin, G.~Verna, H.~van
  Kemenade,
  \href{https://doi.org/10.5281/zenodo.7741289}{cta-observatory/pyirf: v0.8.1 -
  2023-03-16} (Mar. 2023).
\newblock \href {https://doi.org/10.5281/zenodo.7741289}
  {\path{doi:10.5281/zenodo.7741289}}.
\newline\urlprefix\url{https://doi.org/10.5281/zenodo.7741289}

\bibitem{cherenkov_telescope_array_observatory_2016_5163273}
C.~T.~A. Observatory, C.~T.~A. Consortium,
  \href{https://doi.org/10.5281/zenodo.5163273}{{CTAO Instrument Response
  Functions - version prod3b-v2}} (Apr. 2016).
\newblock \href {https://doi.org/10.5281/zenodo.5163273}
  {\path{doi:10.5281/zenodo.5163273}}.
\newline\urlprefix\url{https://doi.org/10.5281/zenodo.5163273}

\bibitem{cherenkov_telescope_array_observatory_2021_5499840}
C.~T.~A. Observatory, C.~T.~A. Consortium,
  \href{https://doi.org/10.5281/zenodo.5499840}{{CTAO Instrument Response
  Functions - prod5 version v0.1}} (Sep. 2021).
\newblock \href {https://doi.org/10.5281/zenodo.5499840}
  {\path{doi:10.5281/zenodo.5499840}}.
\newline\urlprefix\url{https://doi.org/10.5281/zenodo.5499840}

\bibitem{Breiman1984}
L.~Breiman, J.~H. Friedman, R.~A. Olshen, C.~J. Stone, Classification And
  Regression Trees, 1st Edition, Chapman and Hall/CRC, New York, 1984.

\bibitem{Veen2021}
R.~van Veen, M.~Biehl, G.-J. de~Vries, sklvq: Scikit learning vector
  quantization, Journal of Machine Learning Research 22~(231) (2021) 1--6.

\end{thebibliography}





\end{document}